\documentclass[twocolumn,aps,prc,showpacs,superscriptaddress,preprintnumbers,floatfix,nofootinbib]{revtex4}
\usepackage{epsfig,graphics}
\usepackage{graphicx}% Include figure files
\usepackage{dcolumn}% Align table columns on decimal point
\usepackage{bm}% bold math
\usepackage{amsmath}
\usepackage[usenames]{color}
\usepackage{ulem} %% for strike-through

\newcommand{ \pt } {${p_{T}}$}

\newcommand{ \rcp } {${R_{cp}}$}

\usepackage{CJK}
\usepackage{epstopdf}
\hfuzz=\maxdimen
\begin{document}
%\begin{CJK*}{GBK}{song}

\title{Nuclear medium effect on nuclear modification factor of protons and pions in intermediate-energy heavy ion collisions}
\author{M. Lv}
\affiliation{Shanghai Institute of Applied Physics, Chinese Academy of Sciences, Shanghai 201800, China}
\affiliation{University of Chinese Academy of Sciences, Beijing 100049, China}
\author{Y. G. Ma}
\thanks{Corresponding author, Email: ygma@sinap.ac.cn}
\affiliation{Shanghai Institute of Applied Physics, Chinese Academy of Sciences, Shanghai 201800, China}
\affiliation{ShanghaiTech University, Shanghai 200031, China}
\author{J. H. Chen}
\author{D. Q. Fang}
\affiliation{Shanghai Institute of Applied Physics, Chinese Academy of Sciences, Shanghai 201800, China}
\author{G. Q. Zhang}
\affiliation{Shanghai Institute of Applied Physics, Chinese Academy of Sciences, Shanghai 201800, China}

\date{\today}

\begin{abstract}

Nuclear modification factor ($R_{cp}$) of protons and pions are investigated by simulating Au + Au collisions from 0.8 to 1.8$A$ GeV in a framework of an isospin-dependent quantum molecular dynamics (IQMD) model. $R_{cp}$ of protons rises with the increase of \pt~ at different beam energies owing to radial flow and Cronin effect.
The rate of increase of \rcp~ is suppressed at higher beam energies. The significant difference of $R_{cp}$ between protons and pions indicates different medium effects between protons and pions. By changing the in-medium nucleon-nucleon cross section, the $R_{cp}$ of protons changes a lot, while the $R_{cp}$ of pions does not. Taking the pion absorption into account, the $R_{cp}$ of pions becomes close to unity without $p_{T}$ dependence after deactivating the reaction $\pi N \rightarrow \Delta$, while there is nearly no change on proton. This suggests that the pion absorption plays a dominant role on pion dynamics and have slight effect for proton dynamics.
\end{abstract}

\maketitle

\section{Introduction}

  One of the main goals of research in the intermediate energy heavy ion collisions (HICs) has focused on learning
  the bulk properties of hot and compressed nuclear matter and its transport mechanisms since the last thirty
  years \cite{Danil,Archelin_QMD,Hartnack_IQMD}. Transport models such as BUU-type \cite{Bertsch_BUU,Cassing_BUU,Bona} and Molecular Dynamics type \cite{Archelin_QMD,Hartnack_IQMD,FMD,AMD,EQMD}  have been successful in describing the reaction dynamics of
  the low and intermediate energy heavy-ion collisions. The two main ingredients of the nuclear transport process are
  the nucleonic mean field and nucleon-nucleon (NN) binary interaction.
  Recently, the medium effects on nucleon-nucleon cross section (NNCS) have been widely investigated by replacing the NNCS in vacuum with an
  in-medium one \cite{Ter,GQLi,XZCai,JYLiu,YXZhang,Chen}.
 Since a high density region for the compressed nuclear matter could be reached up to 2-3 times normal nuclear matter density $\rho_{0}$
  before it expands during the process of heavy-ion collision at 1-2$A$ GeV, the in-medium NNCS is therefore an important component in these phenomenological simulations due to its close relation with the density.

  The nuclear modification factor (NMF) has been extensively studied in relativistic heavy ion collisions
 in recent years \cite{RHIC_NMF1,RHIC_NMF2,RHIC_NMF3,RHIC_NMF4,STAR_NMF,STAR_NMF2}. In these studies, unanimous results have demonstrated that the NMF
 is suppressed at high $p_{T}$  owing to partonic energy loss effect. However, the $R_{cp}$ of protons show a  rise with $p_{T}$
 at the low and moderate $p_{T}$ range  in intermediate energy HICs, which was argued to be an indication of combined effect from radial flow and Cronin effect \cite{MLyu}. Nuclear modification factor of light nuclei has been also investigated for the first time in a framework of thermal and coalescence models and a number of constituent quark scaling like behavior of pions and protons was exhibited \cite{Ma-NMF}. On the other hand, the nuclear stopping can provide the information on the nuclear equation of state (EoS), in medium nucleon-nucleon cross section as well as the degree of equilibrium \cite{Andronic,GQZhang}. Furthermore, the magnitude of nuclear stopping may have a direct relationship with the enhancement of $R_{cp}$. Besides, some other physical quantities, such as radial flow, temperature and viscosity, can also provide abundant information about dense hadronic matter formed in heavy ion collisions \cite{Reisdorf,CMuntz,CLZhou,XuJ}.

  In the present work, the nuclear modification factors $R_{cp}$ of protons and pions at different incident energies
are investigated systematically. The nuclear medium effect from in-medium NNCS has been studied, while
  the pion absorption effect has also been discussed.

  The article is organized as following: In Sec.II a brief introduction on IQMD model is described. Sec. III gives a description of pion dynamics in the IQMD model from which  the pion spectra are compared with the FOPI experimental results and a preferable matching is obtained.
  The $R_{cp}$ of protons and pions versus $p_{T}$ and the radial flow are calculated at different incident energy in Sec. IV.
  Then we turn to the study of in-medium nucleon-nucleon cross section which indicates the effective nucleon-nucleon interaction  in the dense matter environment in Sec. V. At last, the effect of pion absorption has been investigated in Sec. VI.
  Summary is  given in the last section.

\section{Brief description of IQMD model}

The Quantum Molecular Dynamics model is a transport model which is based on a many body theory to describe  heavy ion collisions from intermediate to relativistic energy~\cite{Archelin_QMD,Archelin_MDQMD,Hartnack_QMD,Hartnack_IQMD}. An extended version, so-called the Isospin dependent Quantum Molecular model (IQMD) which considers the isospin effects is suitable to investigate asymmetric nuclear system.  IQMD can successfully treat collective flow, multi-fragmentation, isospin effects, transport coefficients, giant resonance, and strangeness production etc \cite{Rev1,CLZhou,Kumar,NST,NST2}.

Wave function of each nucleon in the IQMD model is described as a coherent state with the form of Gaussian wave packet,
       \begin{equation}
       \phi_i(\vec{r},t)= \frac{1}{(2\pi L)^{3/4}}exp(-\frac{(\vec r-\vec r_i(t))^{2}}{4L})exp(\frac{i\vec r\cdot \vec p_i(t)}{\hbar}),
       \label{eq1}
       \end{equation}
where  ${\vec r_i}$ and ${\vec p_i}$ are the time dependent variables which describe the center of the packet in coordinate and momentum space, respectively. The parameter $L$, related to the width of  wave packet in coordinate space, is determined by the size of reaction system. Usually $L$ = 2.16 fm$^{2}$ for Au+Au system. The wave function of the system is the direct product of all the nucleon wave functions without considering the Fermion property of nucleon:
     \begin{equation}
      \Phi(\vec{r},t) = \prod_i \phi_i (\vec{r},t).
      \label{eq1.5}
      \end{equation}
As a compensation, Pauli blocking is employed in the initializations and collision process to restore some parts of the quantum property of many Fermion system.

By applying a generalized variational principle on the action of the many-body system, one can get the  equations of motion for ${\vec p_i}$ and ${\vec r_i}$, which are listed as follows
       \begin{equation}
       \vec p_i=-\frac{\partial \left\langle H \right\rangle}{\partial \vec r_i};\\
       \vec r_i=\frac{\partial \left\langle H \right\rangle}{\partial \vec p_i}.
       \label{eq2}
       \end{equation}

The Hamiltonian $\left\langle H \right\rangle=\left\langle T \right\rangle+\left\langle V \right\rangle$ where $T$ is the kinetic energy, the potential $V$ is expressed by
\begin{equation}
\label{meanfield}
\langle V \rangle = \frac{1}{2} \sum_{i} \sum_{j \neq i}
 \int f_i(\vec{r},\vec{p},t) \,
V^{ij}  f_j(\vec{r}\,',\vec{p}\,',t)\, d\vec{r}\, d\vec{r}\,'
d\vec{p}\, d\vec{p}\,',
\end{equation}
where the Wigner distribution function $f_i
(\vec{r},\vec{p},t)$, which is the phase-space density of the $i$th
nucleon, is obtained by applying the Wigner transformation on the
single nucleon wave function:
       \begin{equation}
       f_i(\vec r,\vec p,t)=\frac{1}{(\pi\hbar)^{3}}e^{-(\vec r-\vec r_i(t))^{2}\frac{1}{2L}} e^{-(\vec p-\vec p_i(t))^2\frac{2L}{(\hbar)^{2}}}.
       \label{eq3}
       \end{equation}

The baryon-potential consists of the real part of the G-Matrix which is supplemented by the Coulomb interaction between the charged particles. The former one can be divided into three parts, the Skyrme-type interaction, the finite-range Yukawa potential, and the momentum-dependent interaction (MDI) parts.
The two-body interaction potential $V^{ij}$ in Eq.~\ref{meanfield} can be expressed as follows:
       \begin{equation}
       \begin{split}
       V^{ij}&=G^{ij}+V_{Coul}^{ij}=V_{Skyme}^{ij}+V_{Yuk}^{ij}+V_{MDI}^{ij}+V_{Coul}^{ij}\\
       &=t_{1}\delta(\vec x_{i}-\vec x_{j})+t_{2}\delta(\vec x_{i}-\vec x_{j})\rho^{\gamma-1}(\vec x_{i})\\
       &+t_{3}\frac{exp(-(\vec x_{i}-\vec x_{j})/\mu)}{(\vec x_{i}-\vec x_{j})/\mu}  \\
       &+t_{4}ln^{2}[1+t_{5}(\vec p_{i}-\vec p_{j})^{2}]\delta(\vec x_{i}-\vec x_{j}) \\
       &+\frac{Z_{i}Z_{j}e^{2}}{\vec x_{i}-\vec x_{j}}.
       \label{eq4}
       \end{split}
       \end{equation}
The symmetry potential between protons and neutrons corresponding to the Bethe-Weizsacker mass formula can be taken as
       \begin{equation}
       V_{sym}^{ij}=t_6\frac{1}{\rho_0}T_{3i} T_{3j} \delta(r_i-r_j)
       \label{eq5}
       \end{equation}
with $t_6$=100 MeV. By integrating Skyrme part as well as the momentum dependent part of the two-body interaction and introducing the interaction density,
       \begin{equation}
       \rho_{ij}=\frac{1}{(4\pi L)^{3/2}} \sum_{j\neq i} exp[-\frac{(\vec r_{i}-\vec r_{j})^2}{4L}],
         \label{eq6}
       \end{equation}
one can get the local mean field potential which contains the Skyrme potential and momentum dependent potential
       \begin{equation}
       U=\alpha(\frac{\rho}{\rho_0})+\beta(\frac{\rho}{\rho_0})^{\gamma}+\frac{\rho}{\rho_0} \int d\vec{p}\,'g(\vec{p}\,') \delta\ln^2[\epsilon(\vec{p}-\vec{p}\,')^2+1],
       \label{eq7}
       \end{equation}
where $\rho_0$ is the saturation density at ground state, $g(\vec p,t)=\frac{1}{(\pi\hbar)^{3/2}}\sum_{i}e^{-(\vec p-\vec p_i(t))^2\frac{2L}{(\hbar)^{2}}}$ is the momentum distribution function, the interaction density $\rho=\sum_{ij}\rho_{ij}$, and $\alpha$, $\beta$, and $\gamma$ are the Skyrme parameters, which connect tightly with the EOS of the bulk nuclear matter, as listed in Table~\ref{qmd_par}.

       \begin{table}[htbp]
       \caption{Parameter sets for the nuclear equation of state used in the IQMD model. S and H represent the soft and hard equation of state, respectively, M refers to the inclusion of momentum dependent interaction. This table is adapted  from~\cite{Hartnack_IQMD}.}
       \label{qmd_par}
       \centering
       \begin{tabular}{p{35pt} p{38pt} p{38pt} p{38pt} p{38pt} p{38pt}}
          \hline
          \hline
             & $\alpha$ & $\beta$ & $\gamma$ & $\delta$ & $\epsilon$ \\
             & (MeV)    & (MeV)   &          &  (MeV)   & $(\frac{c^{2}}{(GeV)^{2}})$ \\
          \hline
          S  & -356 & 303 & 1.17 & -    & -   \\
          SM & -319 & 320 & 1.14 & 1.57 & 500 \\
          H  & -124 & 71  & 2.00 & -    & -   \\
          HM & -130 & 59  & 2.09 & 1.57 & 500 \\
          \hline
          \hline
       \end{tabular}
       \end{table}

With the help of coalescence mechanism, the information of fragments produced in HICs can be identified in IQMD. A simple coalescence rule to form a fragment is used with the criteria $\Delta r$ = 3.5 fm and $\Delta p$ = 300 MeV/c between two considered nucleons.

\section{Delta resonance production and decay, and Pion dynamics}

In the present work, the production of pion is considered when inelastic scattering occurs. Pions are produced via the decay of $\Delta$ resonance, the following inelastic reaction channels have been taken into account explicitly at 1$A$ GeV domain \cite{Engel}. And the higher mass resonance N$^*$ channel hasn't been considered owing to its contribution is negligible in this energy domain.

       \begin{equation}
       \begin{array}{l}
       (a)N N \rightarrow N \Delta ~~~(hard ~~ \Delta\ ~~ production), \\
       (b)\Delta\quad \rightarrow N \pi ~~~(\Delta\ ~~ decay), \\
       (c)\Delta N \rightarrow N N ~~~(\Delta\ ~~ absorpion), \\
       (d)N \pi\, \rightarrow \Delta ~~~(soft ~~ \Delta\ ~~ production).
       \label{eq9}
       \end{array}
       \end{equation}

Elastic $\pi-\pi$, $\pi-N$, $\pi-\Delta$, $\Delta-\Delta$ and $\Delta-N$ scattering channels are not taken into account. In the processes  (a) and (d), the experimental cross section and the elastic NN collision are used as shown below.

  The elastic nucleon-nucleon scattering angular distribution is described as $\frac{d\sigma_{el}}{d\Omega} \sim e^{c \cdot t}$,  where $t$ is equal to $-2p^{2}(1-cos(\theta))$, $p$ is the total momentum in the center of mass system (c.m.s) of these two colliding nucleons, the squared momentum transfer $c$ is a function of $\sqrt s$ and defined as,
  \begin{equation}
  c(\sqrt s)=6\cdot\frac{(3.65\cdot(\sqrt{s}-0.18766))^6}{(1+(3.65\cdot(\sqrt{s}-0.18766))^6)},
  \end{equation}
where  $\sqrt{s}$ is the c.m.s energy with GeV unit \cite{Cugnon}.

  The inelastic angular distribution is given by the following function 
  \begin{equation}
  \frac{d\sigma_{el}}{d\Omega} \sim a \cdot e^{b \cdot cos(\theta)} ,
  \end{equation}
 where the parameters $a(\sqrt s)$ and $b(\sqrt s)$ were suggested by Huber and Aichelin and vary in their definition for different intervals of $\sqrt s$. The details can be found in Ref.~ \cite{Huber}.

  Pions produced via $\Delta$ decay propagate with a high thermal velocity under the Coulomb force. The different isospin channels have been considered and the branching ratios use the Clebsch-Gorden coefficient:
       \begin{equation}
       \begin{array}{l}
       (a)~~\Delta^{++} \rightarrow 1(p+\pi^{+})                  \\
       (b)~~\Delta^{+}  \rightarrow 2/3(p+\pi^{0})+1/3(n+\pi^{+}) \\
       (c)~~\Delta^{0}  \rightarrow 2/3(n+\pi^{0})+1/3(p+\pi^{-}) \\
       (d)~~\Delta^{-}  \rightarrow 1(n+\pi^{-})                  \\
       \label{eq10}
       \end{array}
       \end{equation}
 After a $\pi$ is produced, it has two routines, namely $\pi$ absorption $\pi NN \rightarrow \Delta N \rightarrow NN$ and $\pi$ scattering (reabsorption) $\pi N \rightarrow \Delta \rightarrow \pi N$. In this calculation, we named the sum of this two routines   ``$\pi$ absorption process''.

\begin{figure}[htbp]
       \includegraphics[width=0.45\textwidth]{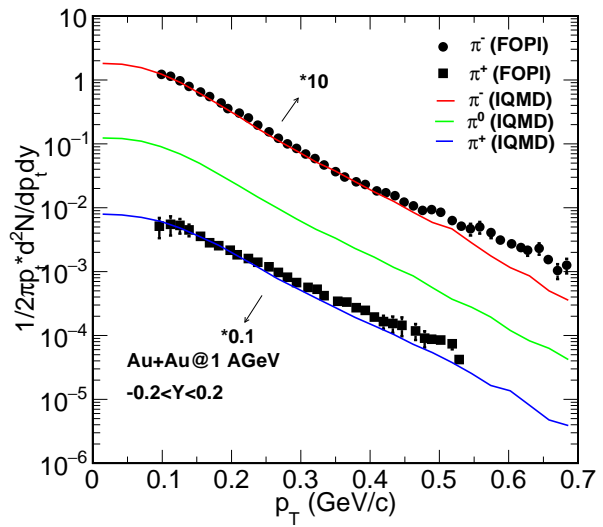}
       \caption{(Color online) Comparison of the invariant $\pi$ production cross section of the measurement and our simulations for mid-rapidity pions in Au + Au collisions at 1$A$ GeV.  The circle represents the FOPI  data of $\pi^{-}$ and the square for $\pi^{+}$, the red, green and blue lines represent different types of $\pi$ from our IQMD simulations.}
       \label{fig:pi_spectraFOPI}
       \end{figure}

In our previous publication, the $p_{T}$ spectra of light charged particles (p, d, t) have been shown to reproduce the experimental  data very well  \cite{MLyu}.  In this work, the $\pi$ spectra from the IQMD model calculation by  a soft equation of state with momentum dependent interaction (SM-EOS) are compared with the FOPI result: it shows  the yields of charged and neutral $\pi$ measured around mid-rapidity ($\pm0.2$) in Au+ Au collisions at 1$A$ GeV by IQMD simulation are comparable  with the FOPI results \cite{Pelte}. Under the ``minimum bias'' condition,  pions are chosen in the rapidity range $-0.2<Y<0.2$, and normalize to dy=1. Figure~\ref{fig:pi_spectraFOPI} shows the spectra can match well with experimental data except a little underestimated in higher $p_{T}$ region because all pions are only produced by $\Delta$ decay in our QMD model calculations.
 
\section{Nuclear modification factor $R_{cp}$ of protons and pions}
\label{sec:rcp-ppi}
  To study nuclear medium effect in Au+Au collisions, it is convenient to introduce a ratio $R_{cp}$ of the particle yield in central collisions to that in peripheral collisions \cite{RHIC_NMF1,Antinori}. Both yields are normalized by corresponding nucleon-nucleon binary collision numbers $\langle N_{coll}\rangle$ (binary scaling),
       \begin{equation}
       R_{cp}=\frac{Yield(central)/{N_{coll}^c}}{Yield(peripheral)/{N_{coll}^p}}.
       \label{eq12}
       \end{equation}
  If nucleus-nucleus collision is a mere superposition of $N_{coll}$ independent nucleon-nucleon collisions, the $R_{cp}$ would be equal to unity. However, owing to the medium effect, such as initial multiple scattering, secondary particle decay and shadowing effect etc, the NMF will show deviation from unity.

In the present work, Au+Au collisions at 0.8, 1.0, 1.2, 1.5 and 1.8$A$ GeV are simulated with IQMD model for the soft equation state with momentum dependent interaction (SM+MDI). The double differential transverse momentum spectra ($\frac{d^{2}N}{2\pi p_{T}dp_{T}dy}$) of protons (pions) at different centralities ($0-20\%, 20-40\%, 40-80\%$) have been obtained with the c.m.s. rapidity cut ($|Y/Y_{proj}|<0.1$, where $Y_{proj}$ is the initial projectile rapidity). All the $R_{cp}$ in this work are obtained by dividing the spectra in the centrality of 0-20\% to the one in the centrality of 40-80\%.

\subsection{$R_{cp}$ of protons and pions at different energies}

\begin{figure}[htbp]
       \includegraphics[width=0.45\textwidth]{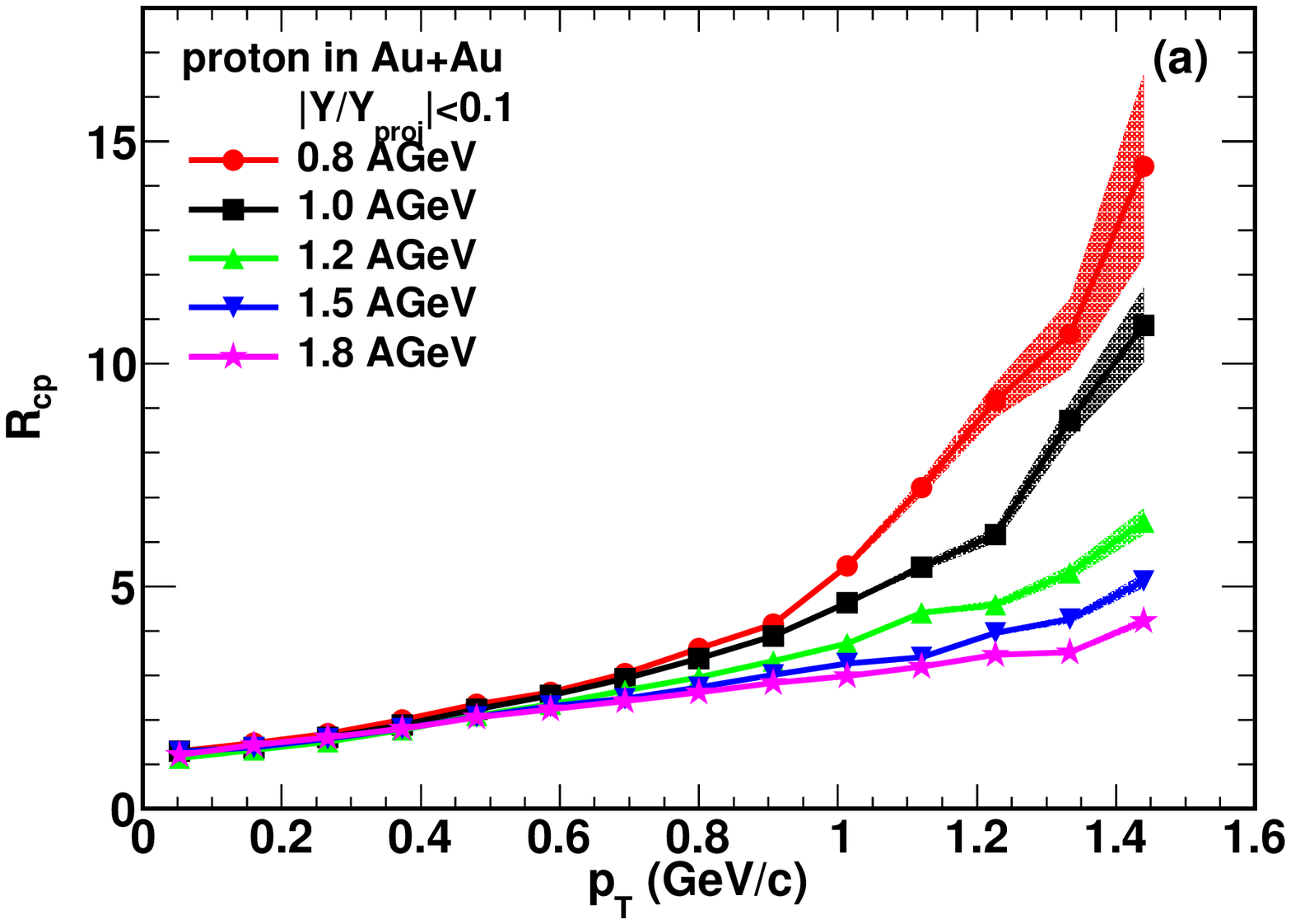}\\
       \includegraphics[width=0.45\textwidth]{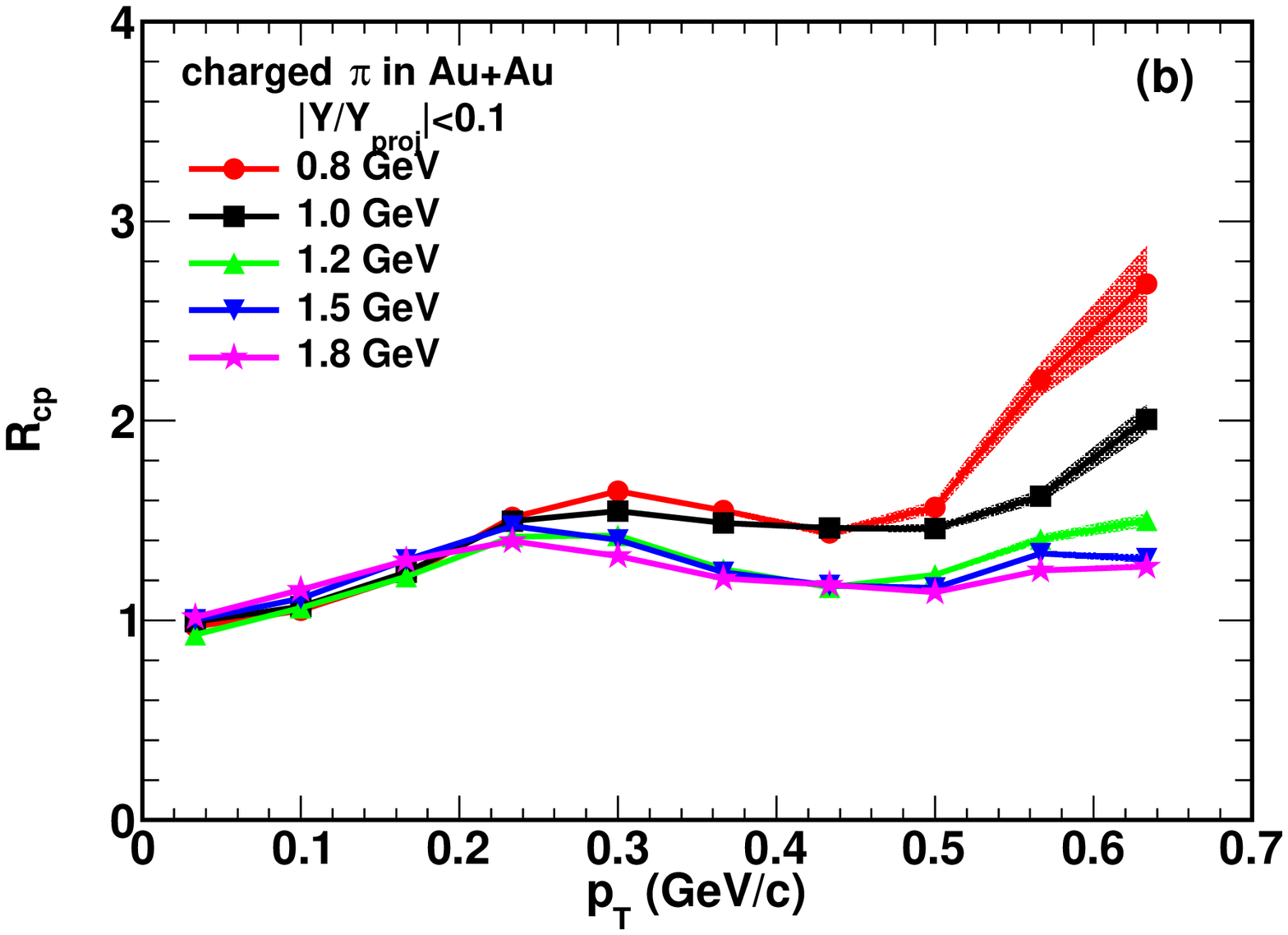}\\
       \caption{(Color online) $R_{cp}$ of protons (a) and pions (b) versus $p_{T}$ at different beam energies. Shadow in  lines represent the statistical error. }
       \label{Rcp_E}
     \end{figure}

Figure~\ref{Rcp_E} shows the $R_{cp}$ of protons and pions for energies 0.8-1.8 AGeV. It is seen that the $R_{cp}$ of protons enhances quickly as $p_{T}$ increases at different beam energies in Fig. \ref{Rcp_E}(a), while the $R_{cp}$ of $\pi$ increases at low $p_{T}$ and level off at high $p_{T}$ as shown in Fig.\ref{Rcp_E}(b). For protons, the strength of $R_{cp}$ enhancement is suppressed at high $p_{T}$ with the increasing of  beam energy.
We noticed that this energy dependence of $R_{cp}$ is quite similar to the preliminary results of Rcp \cite{STAR-Rcp}
which show a monotonic evolution with collision energy from enhancement at low energy due to the Cronin effect to suppression at high energy due to the partonic energy loss when jets pass through the hot dense quark-gluon matter.
On one hand, in the low $p_{T}$ region, radial flow plays a major role in central collisions, which pushes protons to higher $p_{T}$ region and results in the smaller $R_{cp}$ at low $p_{T}$. On the other hand, in the high $p_{T}$ region, the Cronin effect due to multiple nucleon-nucleon scattering effect ~\cite{Antreasyan,Rezaeian} tends to transform the longitudinal momentum into the transverse momentum and the effect becomes stronger with the increasing of $p_{T}$ in central HICs, which leads to larger $R_{cp}$ at high $p_{T}$.
For the pion case, however, the $R_{cp}$ of pions has very different dynamical origin because  they are produced via hard $\Delta$ decay in NN collision and absorbed by nucleon with a large probability (81\% in Au+Au at 1$A$GeV),  \cite{Bass1995}. These differences mean the different nuclear medium effects for protons and pions.

\subsection{Radial flow}

Two important physical quantities, nuclear stopping and the radial flow, which are sensitive to the properties of nuclear bulk matter, have been extensively investigated in a wide range of incident energy from tens of MeV/nucleon to hundreds of GeV/nucleon in many experimental or theoretical work \cite{GQZhang,FFu,Lisa,Herrmann}. The nuclear stopping can be described with a ratio of transverse to parallel quantities (energy or momentum), it reflects  how much energy of original longitudinal motion is transferred into the internal degrees of freedom.

In the compression stage, with nucleon-nucleon r frequently occurring, a highly dense and thermal  nuclear matter zone is formed. New species of particles (such as $\pi$) can be created in the bulk matter with unique condition, and their production yield or emitting pattern can be used to probe the global properties of their surrounding, just as the $R_{cp}$ of pion in the previous Sec.~\ref{sec:rcp-ppi}(A), and pion absorption effect in Sec.~\ref{sec:pi-absorb}.

  And then, the expansion stage occurs owing to the high pressure in the compressed region. While the particles on the surface of coupling matter zone emit outward, and inner particles frequently interact with each other, which cause that the probability of outward motion is larger than the one of inward motion, and then, the collective motion of radial flow grows until the nuclear system freezes out. At high incident energy, hydrodynamics might be suitable to describe these characteristics. A pioneering theoretical model named the Blast-wave model has been put forward by Siemens and Rasmussen \cite{Siemens}, and lots of work based on it have been carried out \cite{FOPI_rflow1,FOPI_rflow2,MLyu,LvM2}.

     \begin{figure}
       \includegraphics[width=0.45\textwidth]{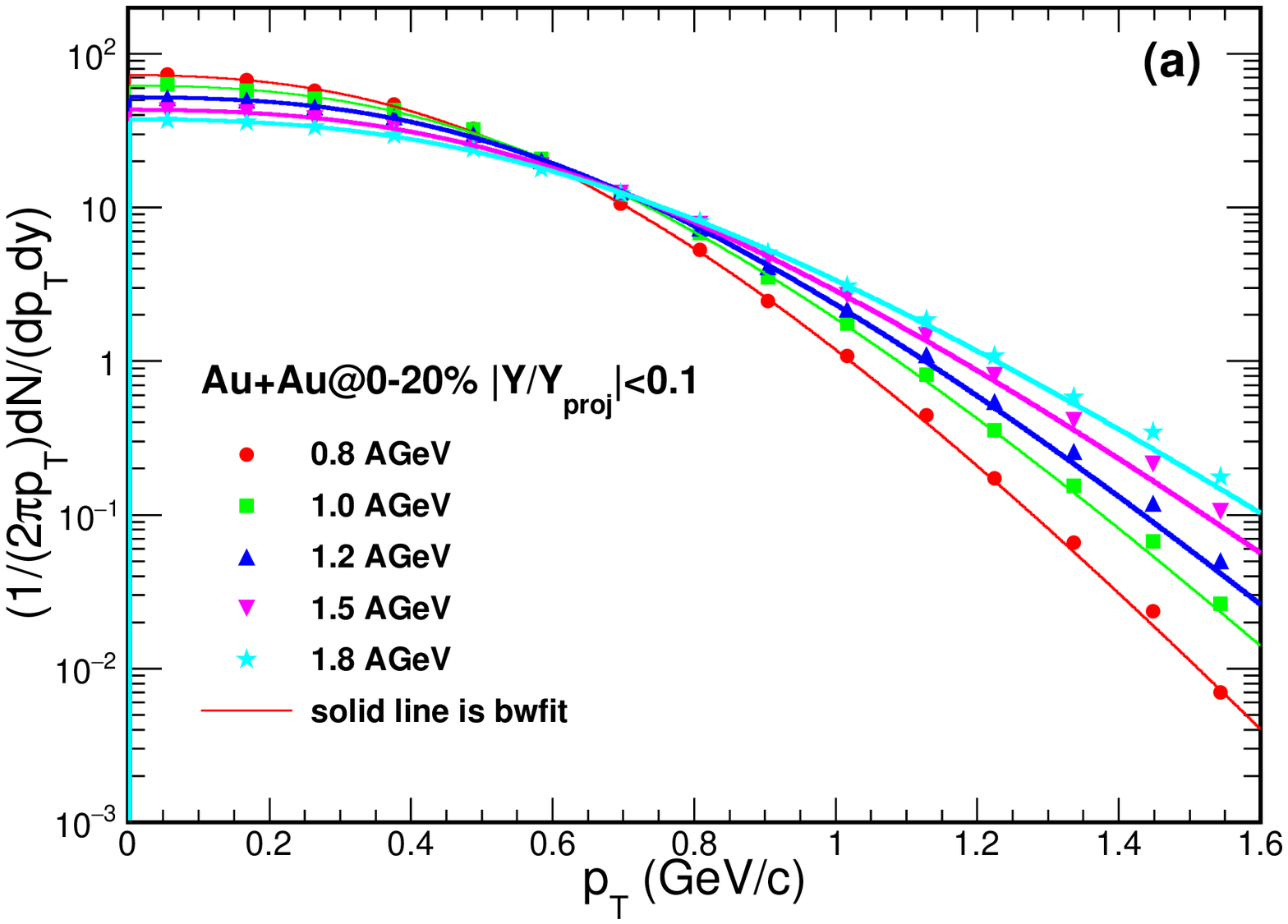}\\
       \includegraphics[width=0.45\textwidth]{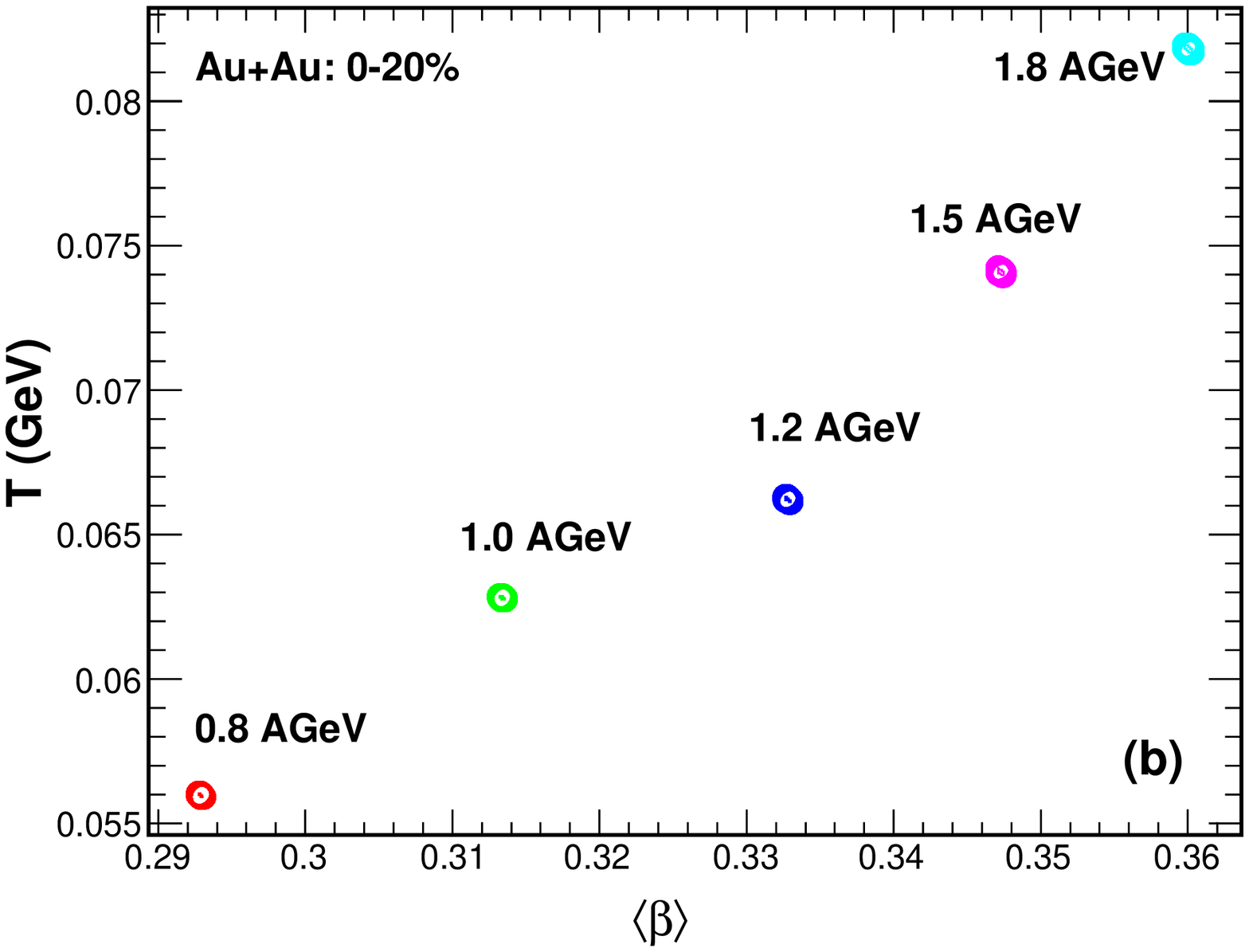}\\
       \caption{(Color online) Blast-wave fits to the spectra and their parameters. Upper pannel: $p_T$ spectra in Au+Au central collisions at 0.8, 1.0, 1.2, 1.5, 1.8$A$ GeV, the solid lines are the Blast-wave fits; Lower pannel:  the Blast-wave fitting contours at different beam energies.}
       \label{fig:rflow}
     \end{figure}

The transverse velocity distribution $\beta_{r}$ in the region of $0 \sim R_{max}$ is described by a self-similar profile, which is parameterized by the surface velocity $\beta_{s}$: $\beta_{r}(r) = \beta_{s}(\frac{r}{R})^\alpha$, where $R_{max}$ is the freeze-out radius, defined as the maximum radius of the expandirng source at thermal freeze-out time, and the $\beta_{S}$ is the particles' radial velocity at the maximum surface where radius is equal to the freeze-out radius, and the exponent $\alpha$ represents the radial flow profile, which describes the evolution of the flow velocity with radius (if $\alpha$=0, it means the uniform velocity;  if $\alpha$=1, it is similar to Habble's law; and if $\alpha$=2, it is hydrodynamical evolution).

Particle spectra are a superposition of individual thermal sources with different $r$,
each boosting with the boost angle $\rho=tanh^{-1}\beta_{r}(r)$ \cite{Schnedermann,MLyu}:
       \begin{equation}
       \frac{dn}{p_{T}dp_{T}}\propto \int_0^{R_{max}} rdr m_{T}I_{0}(\frac{p_{T}sinh\rho}{T_{f}})
       K_{1}(\frac{m_{T}cosh\rho}{T_{f}}),
       \label{eq11}
       \end{equation}
where $K_{1},I_{0}$ are the modified Bessel functions and $T_{f}$ is the freeze-out temperature. The shape of spectra  is essentially determined by the freeze-out temperature, the velocity of the transverse expansion, the flow profile and the mass of the particle.The average flow velocity is estimated by taking an average over the transverse geometry: $\left\langle\beta_{r}\right\rangle=\beta_{S}\frac{2}{2+\alpha}$.

Figures \ref{fig:rflow}(a) shows the Blast-wave fitting on the $p_{T}$ spectra of the Au+Au central collisions at different incident energies, namely 0.8, 1.0, 1.2, 1.5 and 1.8 $A$ GeV. And Fig.~\ref{fig:rflow}(b) are the corresponding fit parameters. From the figure, one can see that both the radial flow $\beta$ and the freeze-out temperature $T$, are increase with beam energy.

\section{In-medium cross section}

     \begin{figure}[htbp]
      \includegraphics[width=0.45\textwidth]{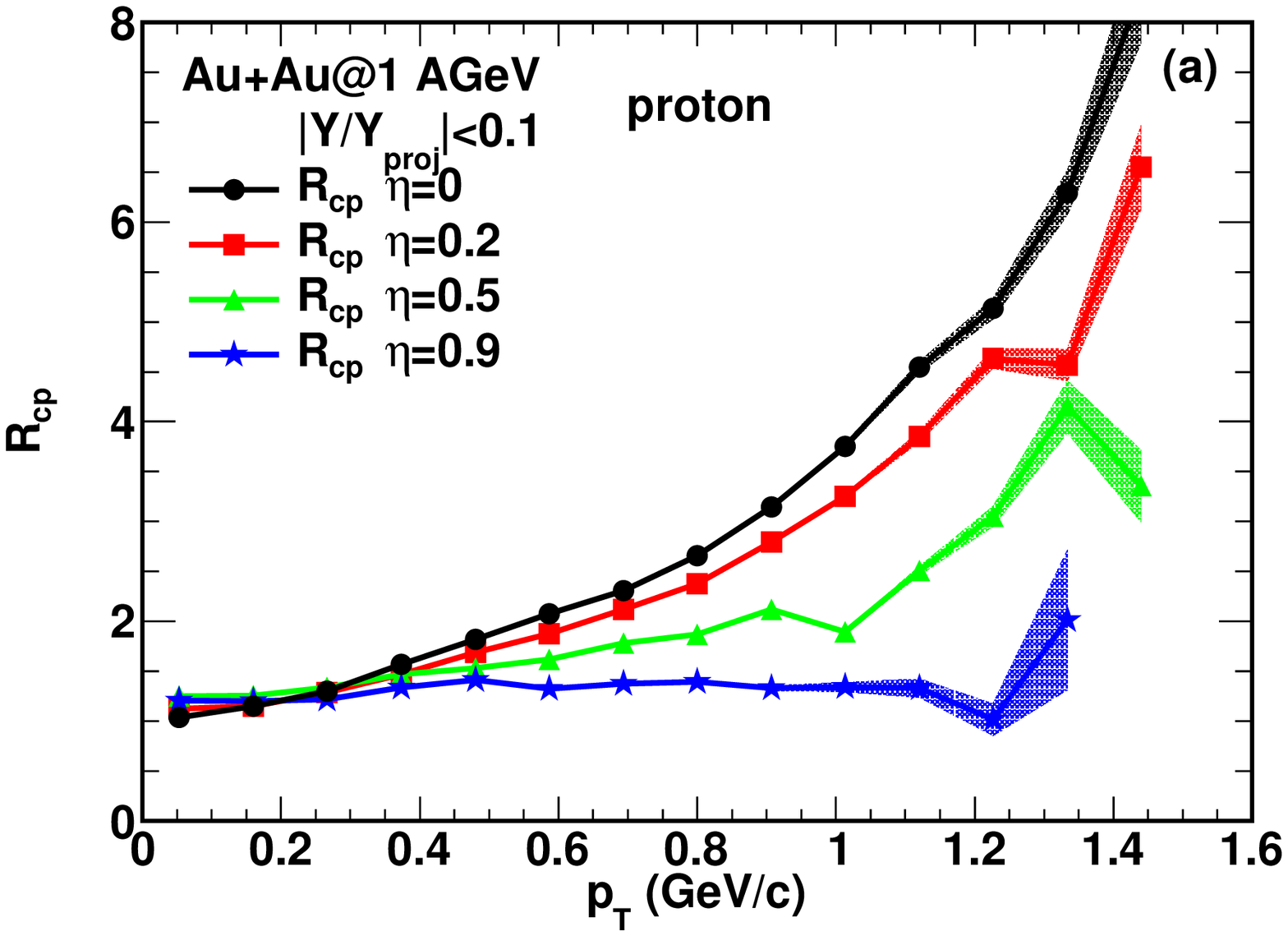}\\
 \includegraphics[width=0.45\textwidth]{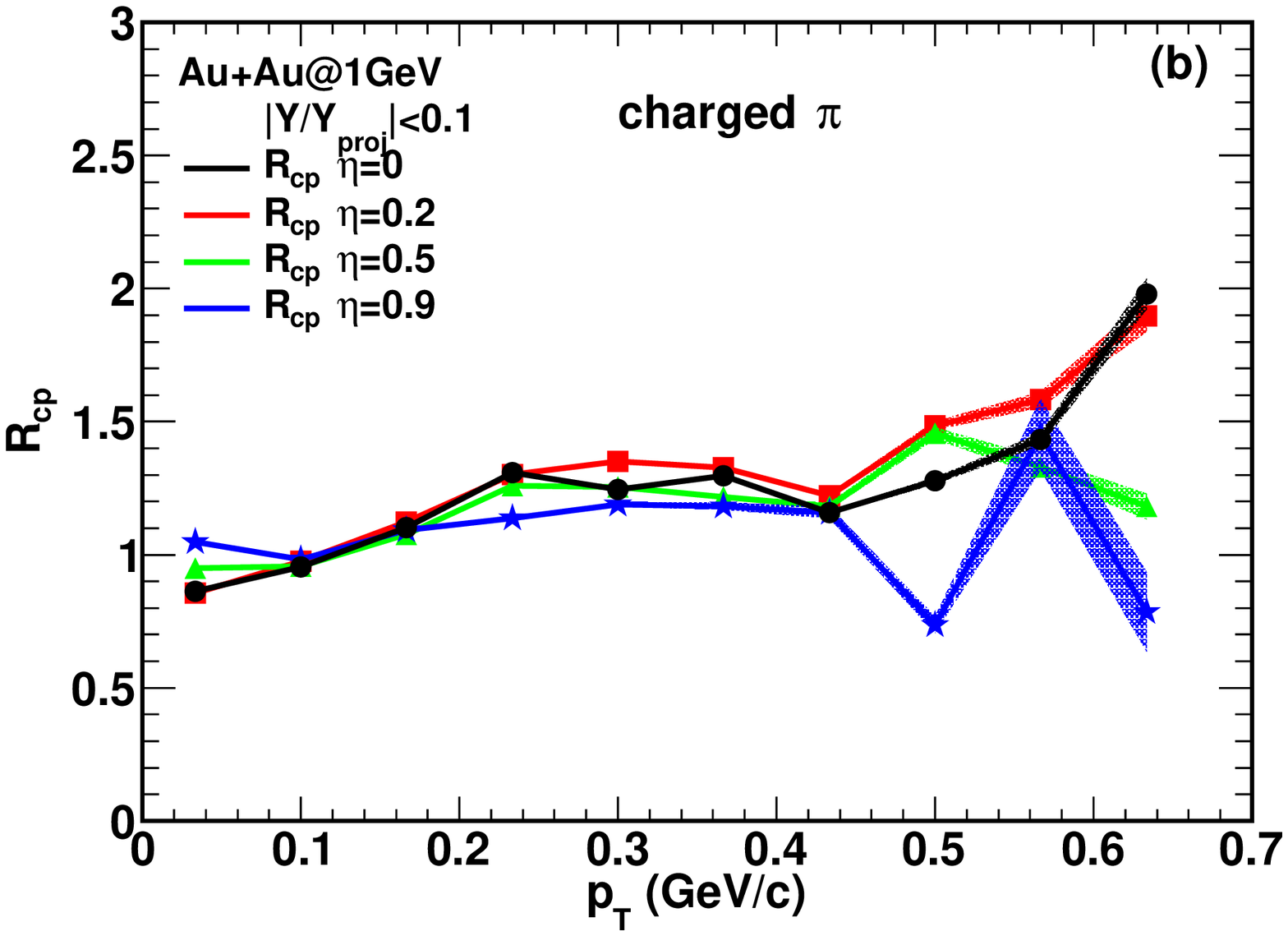}\\
        \caption{(Color online) Dependence of $R_{cp}$ of protons (a) and pions (b) on the reduction factor $\eta$ of in-medium NNCS:
$\eta$=0 (black line), $\eta$=0.2 (red line), $\eta$=0.5 (green line) and $\eta$=0.9 (blue line)}
       \label{nncs_rcp}
     \end{figure}

Usually the free-space nucleon-nucleon cross section $\sigma_{NN}^{free}$ obtained by experimental measurement is used as a default NNCS in QMD model. However, the real in-medium nucleon-nucleon cross section $\sigma_{NN}^{in-medium}$ is different from the free-space nucleon-nucleon cross section because of the effects of Pauli blocking and finite system of nucleus in heavy-ion reactions etc. In-medium two-body cross sections are therefore an indispensable component to compensate the nuclear equation of state  in the QMD simulation \cite{GQLi,XZCai,YXZhang}.
The in-medium nucleon-nucleon cross section can be parameterized from the Particle Data Group with medium modification which can be implemented according to the density dependent prescription \cite{Daffin}:
     \begin{equation}
       \sigma_{NN}=\sigma_{NN}^{free}\cdot(1-\eta\frac{\rho}{\rho_0}),
       \label{eq13}
       \end{equation}
where $\eta$ is the in-medium NNCS reduction factor, varied between 0 and 1, $\rho_{0}$ is the normal nuclear matter density, and  $\rho$ is the local density. Otherwise, the in-medium NNCS scaled by the effective mass $m^{\star}$, $\sigma_{NN}=\sigma_{NN}^{free} \cdot (m^{\star}/m)^2$, has also been employed in the BUU simulation \cite{BALi2005}. The latter scaling presumes that, for given relative momentum, the matrix elements of interaction are not changed between the free space and medium.

     \begin{figure}[htbp]
       \includegraphics[width=0.45\textwidth]{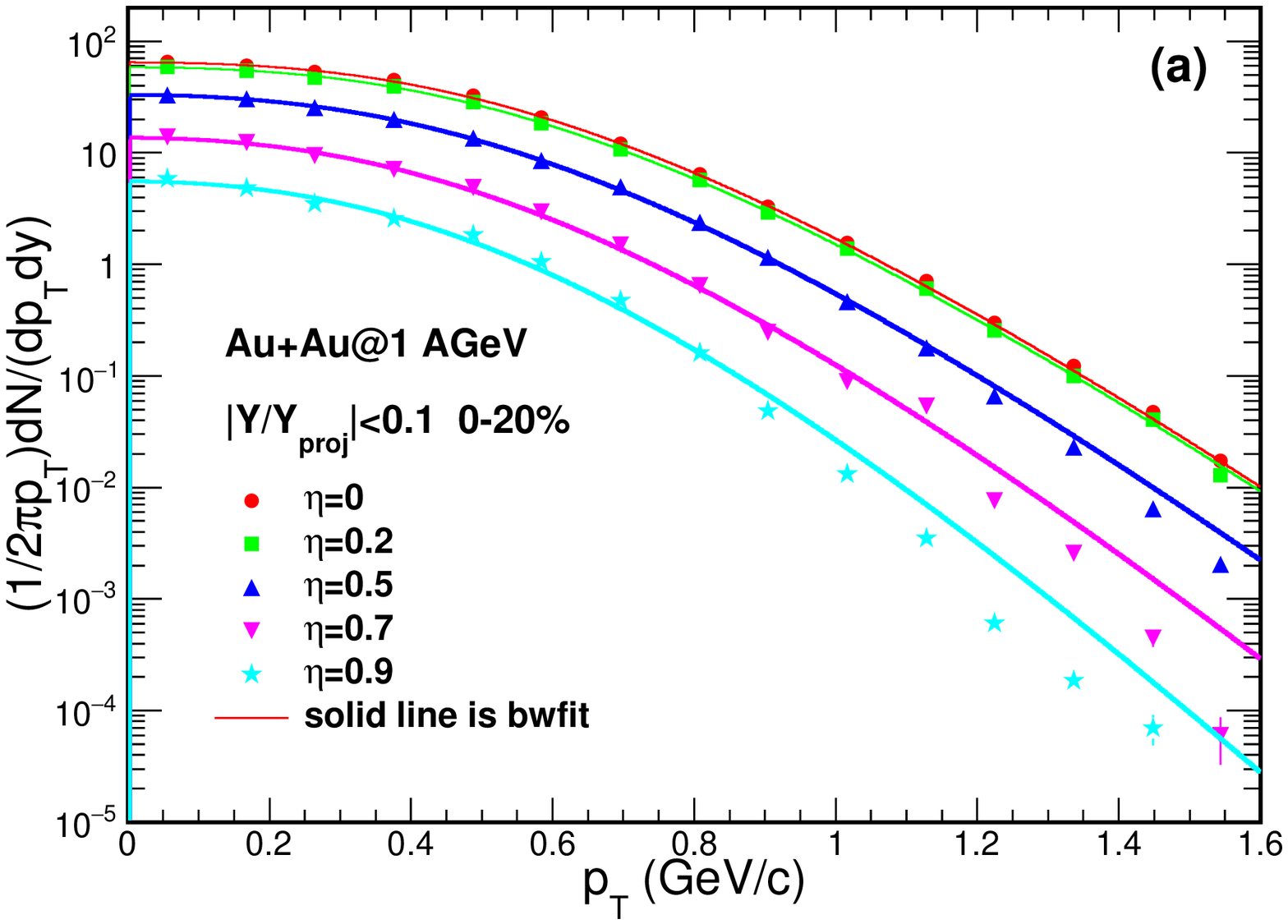}\\
       \includegraphics[width=0.45\textwidth]{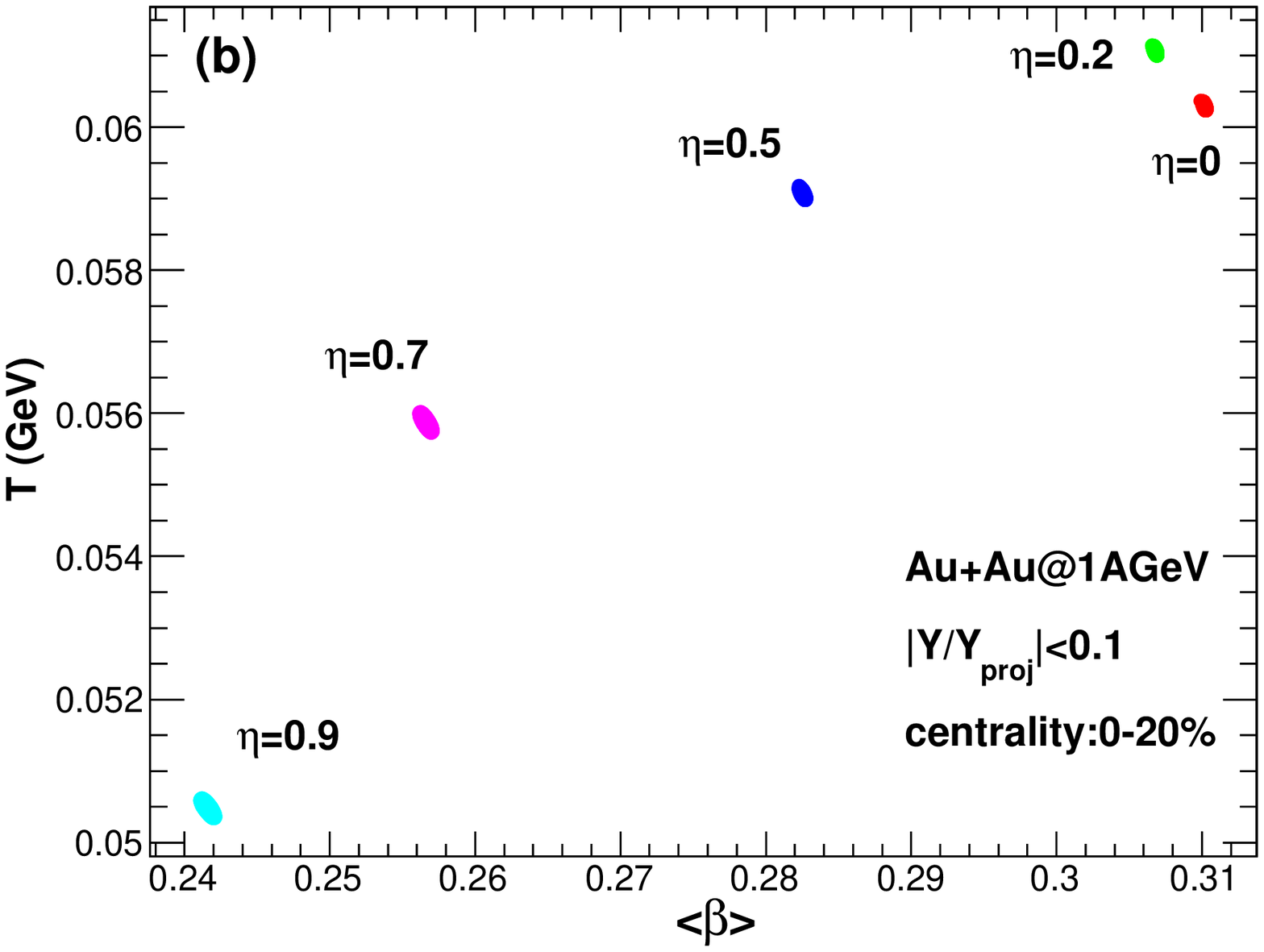}\\
       \caption{(Color online) The  Blast-wave fitting to  $p_{T}$ spectra of protons (upper panel) and their fitting parameters of $\beta$ and $T$ (lower panel)
with  different in-medium NNCS reduction factor in the centrality of  0-20\%.}
       \label{fig:bw-nncs}
     \end{figure}

      \begin{figure}[htp]
       \includegraphics[width=0.45\textwidth]{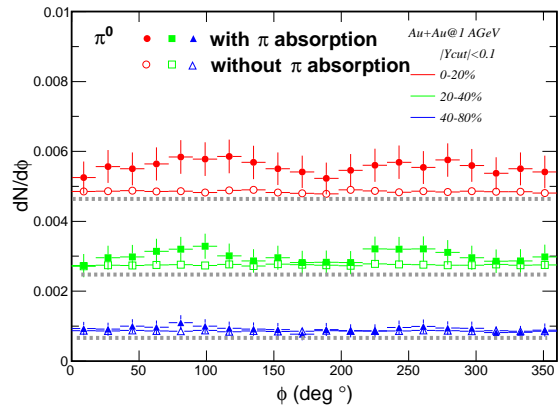}
       \caption{(Color online) Azimuthal distribution of $\pi^{0}$ with or without $\pi$ absorption.}
       \label{fig:pi-azimuthal}
     \end{figure}

In this part, the aim is to draw a conclusion on the in-medium NNCS by investigating $R_{cp}$ with the changing $\eta$ values.
The $\eta$ value at 0.2, 0.5, and 0.9 were used in the simulation of Au+Au at 1$A$ GeV collisions. In one previous work,  the value $\eta$ = 0.2, i.e. 80\% of the free space nucleon-nucleon cross section has been obtained  in \cite{Klakow}. In fact,   the medium effect is different in various ranges of  incident energy and matter density \cite{XZCai}.

\subsection{Effect on $R_{cp}$}

$R_{cp}$ is a good  quantity to study the effect of the in-medium NNCS.  In Fig.~\ref{nncs_rcp}, the $R_{cp}$ of protons (upper panel) and pions (lower panel) are shown with different $\eta$ values.  Due to the limited statistics, there exists fluctuation in high $p_{T}$ region, especially for the $\eta$=0.9 case.The $R_{cp}$ of protons has an increasing trend which is explained by the Cronin effect \cite{MLyu}.
 And its trend is increasing more rapidly with $p_{T}$ in the low $\eta$ value case because of high collision rate between nucleons.
Collisions become certainly less in  higher $\eta$ values and it makes the Cronin effect become less important. On the other hand, the trend of pion's $R_{cp}$ doesn't seem to have any obvious change with different $\eta$ value. The reason might be the similar change of cross section for pion interaction in central and peripheral collisions when the $\eta$ value changes and then lead to the unchanged  $R_{cp}$ with $\eta$. However, the significant decrease of cross section in the $\eta$=0.9 case induces a large fluctuation in high $p_T$ region as shown in Fig.~\ref{nncs_rcp}(b). In the next section,  pion absorption will be discussed in detail.

\subsection{Effect on radial flow}

Bauer {\it et al.}  pointed out that in intermediate energy HIC, nuclear stopping power is determined by both the mean field (EOS) and the in-medium nucleon-nucleon cross section \cite{Bauer}. The nuclear stopping was also proposed as a probe to extract information on the isospin dependence of the in-medium N-N cross section in HIC for the beam energy  from the Fermi energy to about 150$A$ MeV \cite{JYLiu}. A similar physical quantity, radial flow,  is studied in this work. Figure \ref{fig:bw-nncs}(a) shows the $p_{T}$ spectra from the Au + Au central collisions at 1$A$ GeV with different value of the in-medium NNCS factor $\eta$ (0, 0.2, 0.5, 0.7 and 0.9), fitting well with a  function from the Blast-wave model. In Figure \ref{fig:bw-nncs}(b), results demonstrate that the system has larger radial flow and higher temperature when the in-medium NNCS becomes larger (i.e. the lower $\eta$ value). If this scenario is confirmed and with enough precision, then the in-medium NNCS reduction factor might be extracted by  a comparison with the experimental result.

     \begin{figure}[htp]
       \includegraphics[width=0.45\textwidth]{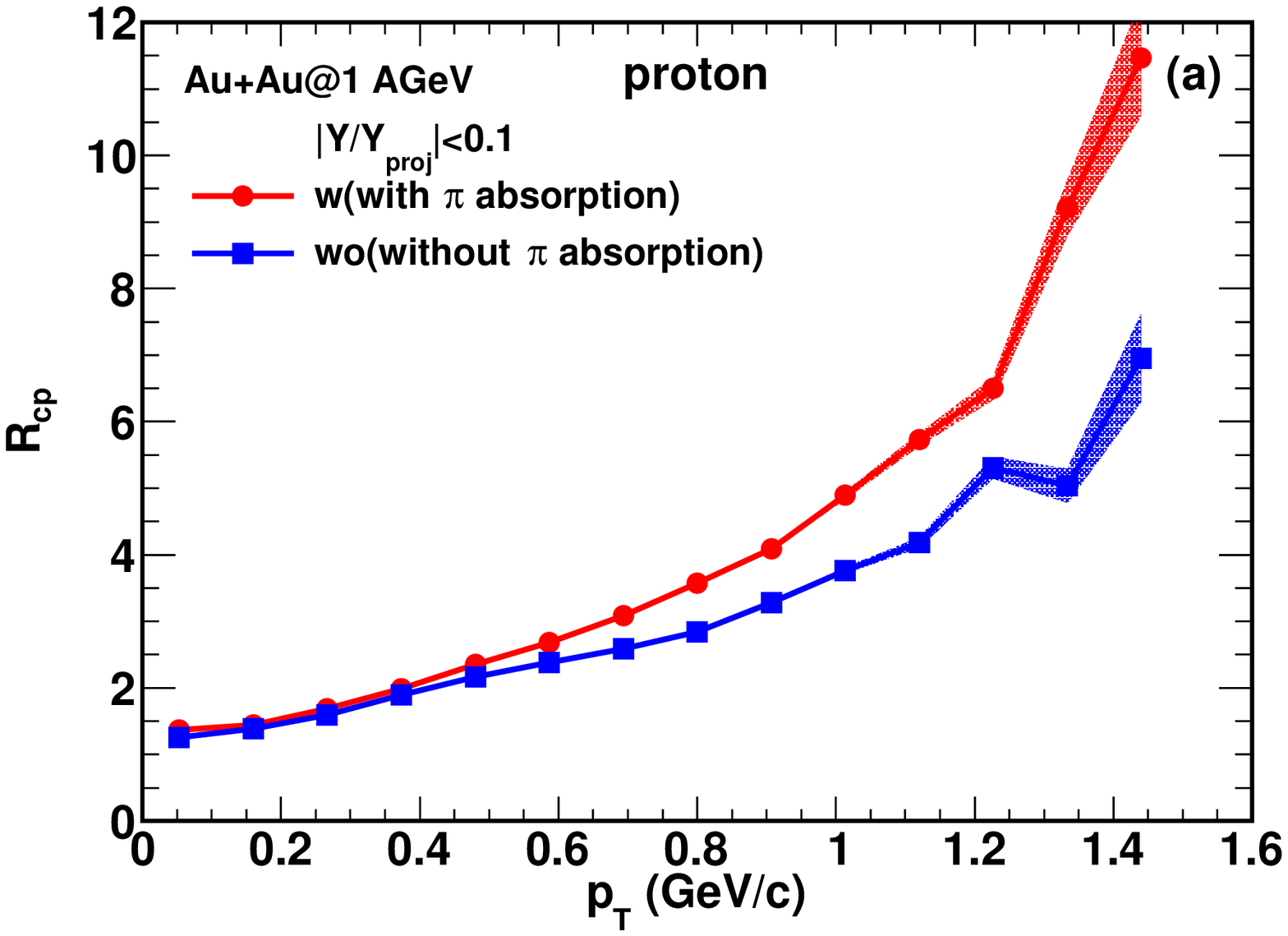}\\
       \includegraphics[width=0.45\textwidth]{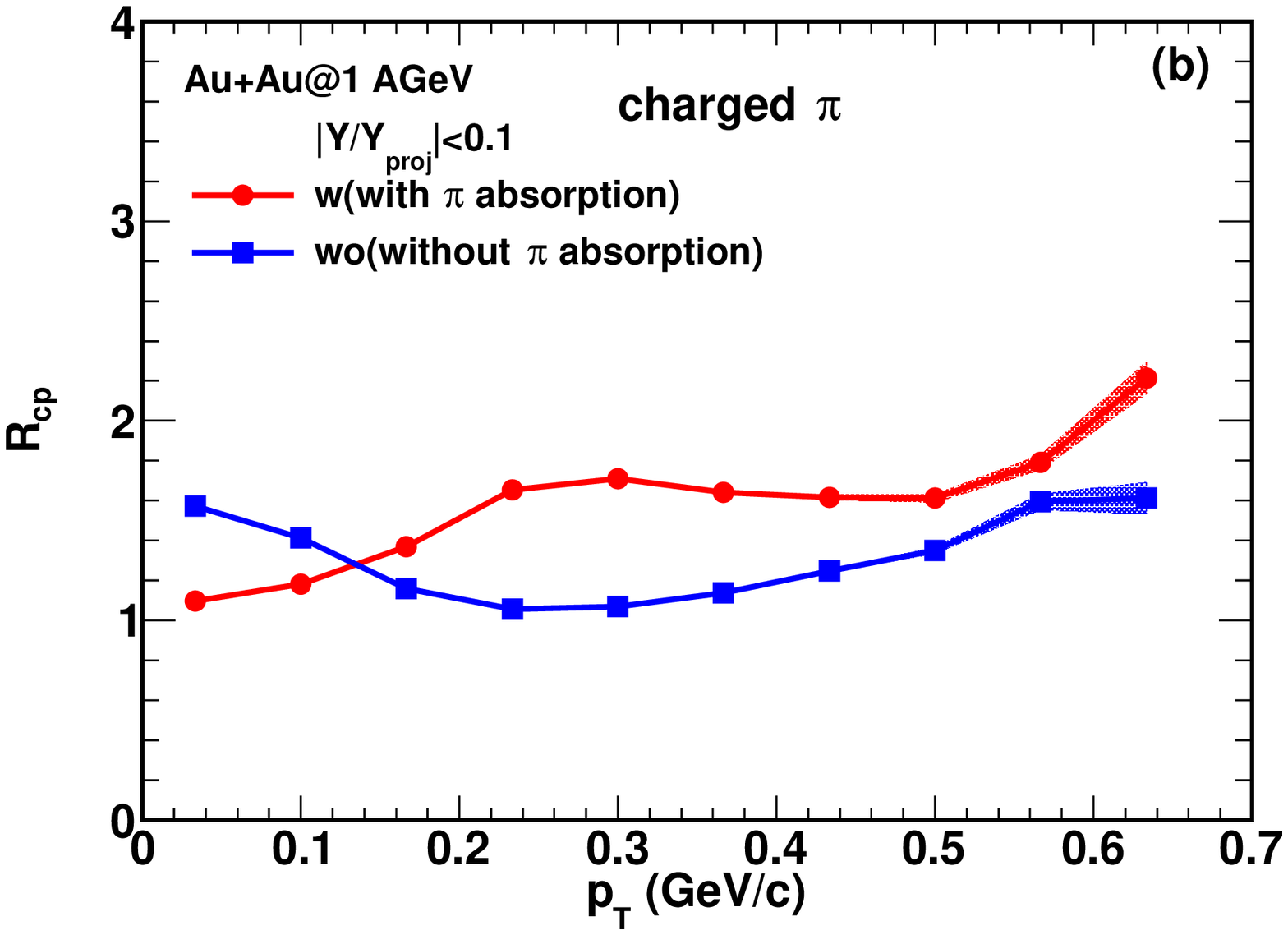}
       \caption{(Color online) $R_{cp}$ of protons and pions versus $p_{T}$ in Au+Au at 1$A$ GeV by the IQMD simulation with soft EoS and MDI. (a) The $R_{cp}$ of protons w/ and w/o $\pi$ absorption channel in IQMD model ; (b) The $R_{cp}$ of the charged $\pi$ w/ and w/o  pion absorption channel.}
       \label{fig:piab-Rcp}
     \end{figure}

\section{ Pion absorption process}
\label{sec:pi-absorb}

 As presented in Sec.II, all the pions are produced by resonance ($\Delta$) decay in this IQMD model as the direct production is very small in intermediate energy range and can be neglected. According to the pion absorption with a large probability, it will  obviously change the phase-space distribution of pions. As discussed in Refs.\cite{Bass1993,BALi1994}, pion scattering channel influences the angular distribution of pions and the pion absorption channel plays an important role for the absolute number of produced pions.  If we turn off this two channels (identify with soft-$\Delta$ production channel ($\pi N \rightarrow \Delta$), here we call ``pion absorption process''), then pion will emit directly from hard-$\Delta$ decay without any scattering and absorption. The hot medium and cold shadowing effects of pions can be quantified by observing $R_{cp}$.
The isotropic angular distribution of $\pi^0$ without pion absorption process is shown in Figure \ref{fig:pi-azimuthal}(a), indicating that the pions are produced originally via hard $\Delta$ decay, while there exists two peaks at 90$^{\circ}$ and 270$^{\circ}$ owing to ``squeezing out'' effect  in the case with pion absorption.

In addition,  $R_{cp}$ of protons and pions are investigated by  a comparison between the cases with and without pion absorption. In Fig.~\ref{fig:piab-Rcp}, the upper panel (a) shows the $R_{cp}$ of proton w/ and w/o pion absorption in Au+Au collisions at incident energy 1$A$ GeV, respectively; and the $R_{cp}$ of changed pion w/ and w/o pion absorption are shown in the bottom panel (b). From this figure, we can see that, while the $R_{cp}$ of protons has a slight difference between the cases w/ and w/o pion absorption, the $R_{cp}$ of pion has been changed prodigiously. And the difference  has demonstrated clearly that ``pion absorption process'' is decisive on pion dynamics.

\section{Summary}

To summarize, nuclear modification factors $R_{cp}$ of protons and pions at different incident energies have been investigated within the IQMD model with SM-EOS. The $R_{cp}$ of protons rises rapidly with the $p_{T}$ increasing at 0.8$A$ GeV owing to radial flow and Cronin effect. And the rising trend becomes slowly with the incident energy increasing. This feature can be explained by the nuclear stopping whose degree decreases with the incident energy increase. On the other hand, the $R_{cp}$ of pion rises slowly at low $p_{T}$ ($<$ 0.3 GeV) and level off at high $p_{T}$($>$ 0.3GeV). It can be understood by these pions being produced from two sources, the hard $\Delta$ decayed pions which emit outward directly and the soft $\Delta$ decay pions which are absorbed and then secondary decayed. The $R_{cp}$ of pions has a little enhancement at low $p_{T}$ because these low energy pions are affected by nucleon dynamics, such as radial flow, and it maintains a saturated trend at high $p_{T}$ owing to no Cronin effect on high energy pions.

The difference of the nuclear modification factor between protons and pions indicates the different interaction mechanisms. Because pion has a large absorption cross section by nucleon, the pions with high transverse momentum  are not enhanced in central collisions, but the protons are. All these observations are consistent with a picture where a dense strongly interacting nuclear matter with a moderate collective flow is most likely formed in central Au + Au collisions over a large rapidity range which results in the strong suppression of charged pion yields at high $p_{T}$ and boosts the protons to higher transverse momentum.

We change the NNCS in nuclear medium between 0.2-0.9. Results demonstrate that radial flow at the central collisions decrease with a smaller in-medium NNCS. The $R_{cp}$ also became weakly increasing with a lower NNCS. Additionally, the $R_{cp}$ obtained in IQMD model without pion absorption is investigated in comparison with the normal case. The results demonstrate that $R_{cp}$ of protons have almost no change after deactivating the reaction channel $\pi N \rightarrow \Delta$, while the $R_{cp}$ of pions have significant difference. This phenomenon reflects that the pion absorption mechanism plays a dominat role on pion dynamics.

This work was supported in part by the Major State Basic Research
Development Program in China under Contract No. 2014CB845401, the
National Natural Science Foundation of China under contract Nos. 11421505,
11220101005, 11322547, 11275250, 11205230.

%\end{CJK*}

\end{document}